\documentclass[aps,prl,twocolumn,superscriptaddress,reprint]{revtex4-2}
\usepackage{graphicx}
\usepackage{amssymb, amsmath}
\usepackage{booktabs}
\usepackage[usenames]{color}
\usepackage{bm}
\usepackage{multirow}
\usepackage{dcolumn}
\usepackage{hyperref}
\usepackage{enumerate}
\usepackage[mathlines]{lineno}
\hypersetup{
    colorlinks=true,
    linkcolor=blue,
    filecolor=gray,      
    urlcolor=blue,
    citecolor=blue,
}

\begin{document}

\title{Neural-network Density Functional Theory Based on Variational Energy Minimization}

\newcommand{\thuphy}{State Key Laboratory of Low Dimensional Quantum Physics and Department of Physics, Tsinghua University, Beijing, 100084, China}
\newcommand{\thuias}{Institute for Advanced Study, Tsinghua University, Beijing 100084, China}
\newcommand{\fscqi}{Frontier Science Center for Quantum Information, Beijing, China}
\newcommand{\riken}{RIKEN Center for Emergent Matter Science (CEMS), Wako, Saitama 351-0198, Japan}

\affiliation{\thuphy}
\affiliation{\thuias}
\affiliation{\fscqi}
\affiliation{\riken}

\author{Yang \surname{Li}}
\thanks{These authors contributed equally to this work.}
\affiliation{\thuphy}

\author{Zechen \surname{Tang}}
\thanks{These authors contributed equally to this work.}
\affiliation{\thuphy}

\author{Zezhou \surname{Chen}}
\thanks{These authors contributed equally to this work.}
\affiliation{\thuphy}

\author{Minghui \surname{Sun}}
\affiliation{\thuphy}

\author{Boheng \surname{Zhao}}
\affiliation{\thuphy}

\author{He \surname{Li}}
\affiliation{\thuphy}
\affiliation{\thuias}

\author{Honggeng \surname{Tao}}
\affiliation{\thuphy}

\author{Zilong \surname{Yuan}}
\affiliation{\thuphy}

\author{Wenhui \surname{Duan}}
\email{duanw@tsinghua.edu.cn}
\affiliation{\thuphy}
\affiliation{\thuias}
\affiliation{\fscqi}

\author{Yong \surname{Xu}}
\email{yongxu@mail.tsinghua.edu.cn}
\affiliation{\thuphy}
\affiliation{\fscqi}
\affiliation{\riken}

\begin{abstract}
Deep-learning density functional theory (DFT) shows great promise to significantly accelerate material discovery and potentially revolutionize materials research. However, current research in this field primarily relies on data-driven supervised learning, making the developments of neural networks and DFT isolated from each other. In this work, we present a theoretical framework of neural-network DFT, which unifies the optimization of neural networks with the variational computation of DFT, enabling physics-informed unsupervised learning. Moreover, we develop a differential DFT code incorporated with deep-learning DFT Hamiltonian, and introduce algorithms of automatic differentiation and backpropagation into DFT, demonstrating the capability of neural-network DFT. The physics-informed neural-network architecture not only surpasses conventional approaches in accuracy and efficiency, but also offers a new paradigm for developing deep-learning DFT methods. 
\end{abstract}

\maketitle

Deep-learning \emph{ab initio} calculation is an emerging interdisciplinary field, which aims to greatly enhance the capability of \emph{ab initio} methods by using state-of-the-art neural-network approaches~\cite{Behler2007,Zhang2018,Schutt2019,Unke2021,Gu2022,deeph2022,deeph-e32023,xdeeph2023,deeph-dfpt2024,deeph-hybrid2023,deeph22024,yu2023efficient,yuan2024equivariant}. For instance, the use of neural-network quantum states significantly improves the accuracy of quantum Monte Carlo calculations~\cite{carleo2017solving, FermiNet}; the integration of deep learning and density functional theory (DFT) can speed up material simulations by several orders of magnitude~\cite{deeph2022,deeph-e32023,xdeeph2023,deeph-dfpt2024,deeph-hybrid2023,deeph22024}. In particular, deep-learning DFT can potentially have a revolutionary impact on future research due to the indispensable role of DFT in various fields ranging from physics and chemistry to materials science. Stimulated by the Materials Genome Initiative launched in 2011, great efforts have been devoted to building computational material databases via DFT. Deep-learning DFT will act as a game changer in the field, since neural-network algorithms can considerably accelerate the construction of bigger material databases, and the bigger data would in turn allow for training more powerful neural-network models. In this context, combining neural networks with DFT database construction holds great promise for advancing materials discovery.

Current research on deep-learning DFT, however, treats the tasks of DFT and neural networks separately: People first compute material datasets by DFT and then train neural-network models by data-driven approaches. With this strategy, individuals can focus on the methodological development of neural networks without needing to delve into the intricacies of DFT algorithms. This results in the development of several valuable neural-network representations of DFT~\cite{deeph2022,deeph-e32023,xdeeph2023,deeph-dfpt2024,deeph-hybrid2023,deeph22024}. In contrast, a more intriguing strategy is to achieve a synergistic combination of neural networks and DFT, termed neural-network DFT, which enables their methodological developments to benefit each other mutually. This objective is theoretically feasible due to the resemblance between the variational principle in physics and the loss minimization rule in deep learning. In analogy to neural-network quantum Monte Carlo~\cite{carleo2017solving,FermiNet}, one may express the total energy as a functional of DFT quantities, such as the Hamiltonian, wave function, and charge density~\cite{Martin2004,li2023dft,mathiasen2024reducing}, and define the energy functional as a loss function for training neural networks. While algorithms for neural networks and DFT are well developed individually, the coherent integration of the two for creating an advanced deep-learning architecture remains elusive.

In this work, we introduce a theoretical framework of neural-network DFT, which unifies the minimization of loss in neural networks with the optimization of the energy functional in DFT. The central idea is to express the total energy as a functional of DFT quantities while simultaneously representing these quantities using neural networks, allowing one to define the energy functional as the loss function of neural networks. We suggest selecting the DFT Hamiltonian as the target quantity to acquire transferable neural-network models and variational energy functionals. To illustrate this concept, we develop a computational code called AI2DFT for implementing neural-network DFT. In AI2DFT, equivariant neural networks are employed to learn the mapping from material structure to DFT Hamiltonian, and DFT algorithms are adapted to be differentiable, introducing modern techniques of automatic differentiation and backpropagation into DFT. Remarkably, neural-network DFT enables physics-informed unsupervised learning, offering superior accuracy and efficiency compared to conventional supervised learning methods. The work establishes a new pathway for developing deep-learning DFT methods.

\begin{center}
\begin{figure}
    \centering
    \includegraphics[width=\linewidth]{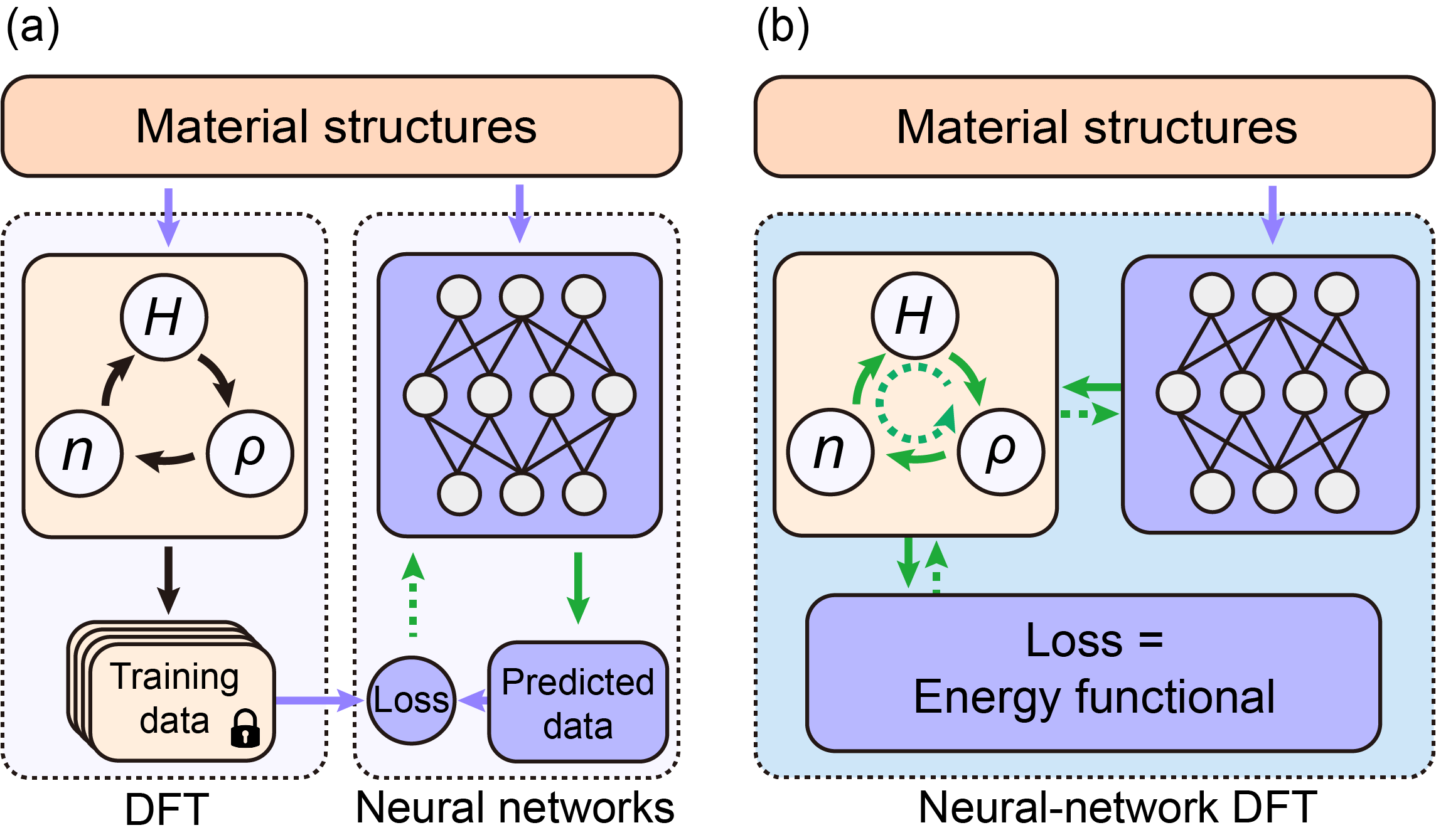}
    \caption{Two different approaches for deep-learning DFT. (a) Conventional data-driven supervised learning, where DFT calculations on varying material structures are performed to generate training data, and the data are used to train neural networks. The DFT Hamiltonian $H$, density matrix $\rho$, and charge density $n$ are iteratively computed. (b) Physics-informed unsupervised learning based on neural-network DFT, which integrates the minimization of the energy functional in DFT with the optimization of the loss function in neural networks. Black arrows depict normal computation, and green arrows denote differential programming computation, with solid and dashed lines representing forward pass and gradient backpropagation, respectively.}
    \label{fig1}
\end{figure} 
\end{center}

The Kohn-Sham DFT is the most widely used \emph{ab initio} approach in material simulations. The method maps the complicated problem of interacting electrons to a simplified problem of noninteracting electrons described by an effective single-particle Kohn-Sham Hamiltonian, which takes the intricate many-body effects into account by employing approximated exchange-correlation functionals~\cite{Martin2004}. Typically, DFT calculations are performed by solving the Kohn-Sham equation via self-consistent field (SCF) iterations. A more fundamental approach involves the variational principle, which computes ground-state properties of materials by minimizing an energy functional of DFT. In fact, the Kohn-Sham equation is formally derived from the variational principle. Although the variational approach is more fundamental and favored by theoretical physics, it is typically not employed for DFT computation. One possible reason is that the variational method requires searching through high-dimensional parameter spaces, which may not be as efficient as solving differential equations iteratively. The situation, however, could potentially change as advanced algorithms and hardware developed for deep learning become available.

The integration of deep learning and DFT has revolutionized the paradigm of method development~\cite{Behler2007,Zhang2018,Schutt2019,Unke2021,Gu2022,deeph2022,deeph-e32023,xdeeph2023,deeph-dfpt2024,deeph-hybrid2023,deeph22024,yu2023efficient}. Nevertheless, previous research in deep-learning DFT has primarily relied on data-driven supervised learning techniques. As illustrated in Fig.\ref{fig1}(a), DFT SCF calculations are first performed to generate training data for varying material structures; neural networks are designed and trained for predicting data resembling DFT results. During this process, the DFT computation and neural-network optimization are separated. We call this scenario ``neural networks and DFT''.

An intriguing strategy is to intimately integrate neural networks and DFT together, termed neural-network DFT [Fig.\ref{fig1}(b)]. This advanced architecture allows us to pursue the synergistic effects between the two. For instance, algorithms for neural networks and DFT may be shared with each other, and their developments can be mutually stimulated. More importantly, by explicitly introducing the knowledge of DFT into deep learning, neural network models might be trained to better emulate real physics than the previous data-training approach.

Variational DFT is preferred for implementing neural-network DFT, because the minimization of energy functional in DFT is similar in spirit to the optimization of loss function in neural networks. The total energy of DFT can be implicitly written as a functional of charge density $n$:
\begin{align}
\label{E_DFT}
E_{\text{DFT}}[n] = T_s + E_{\text{ext}}[n] + E_{\text{Hartree}}[n] + E_{\text{XC}}[n] + E_{II}, \nonumber
\end{align}
which includes the single-particle kinetic energy $T_s$, the external potential energy $E_{\text{ext}}$, the Hartree energy $E_{\text{Hartree}}$, the exchange correlation energy $E_{\text{XC}}$, and the classic interaction between nuclei $E_{II}$~\cite{Martin2004}. Note that $n$ can be derived from other DFT quantities, such as the Kohn-Sham eigenstates ($\{ \psi_i \}$), density matrix ($\rho$), and DFT Hamiltonian ($H_{\text{DFT}}$). The subscript ``DFT'' will be omitted without leading to confusion. Hence different kinds of energy functionals can be defined: $E[Q]$~\cite{Martin2004}, where the target quantity $Q = n, \{ \psi_i \}, \rho$, or $H$. Inspired by neural-network quantum Monte Carlo methods~\cite{carleo2017solving}, neural networks are utilized to represent the target DFT quantity, denoted as $Q_{\theta}$, where ${\theta}$ represents parameters or weights of neural networks. Consequently, the total energy becomes a function of neural-network parameters, denoted as $E[Q_{\theta}]$. In principle, one may use  $E[Q_{\theta}]$ as the loss function, and the optimization of neural networks naturally completes the computation of variational DFT, as depicted in Fig.~\ref{fig1}(b).

The selection of target DFT quantity $Q$ merits careful consideration. Here, we present three selection criteria. Firstly, the target of deep learning should closely adhere to the ``nearsightedness'' or locality principle proposed by Walter Kohn~\cite{Kohn1996,Prodan2005}, thereby enhancing the transferability of neural network models. Local physical quantities such as $\rho$, $n$, and $H$ meet the criterion~\cite{Martin2004,deeph2022}, whereas nonlocal quantities such as Kohn-Sham eigenstates do not. Secondly, neural network representations that satisfy the fundamental requirements for the target quantity should be readily available. Here are some essential requirements: $n$ is positive in value and normalized to a constant; $H$ is Hermitian; $\rho$ at zero temperature is idempotent~\cite{Martin2004, Kohn1996}. While the former two conditions are relatively straightforward to achieve in deep learning, meeting the strict idempotency requirement of $\rho$ remains challenging. Thus, we may further consider $E[H]$ or $E[n]$. Thirdly, the energy functional should be variational. $E[H]$ satisfies the criterion~\cite{Martin2004}, whereas $E[n]$ does not~\cite{HWF_H}, meaning that minimizing $E[n]$ may not yield the correct ground state. Based on the above considerations, we choose the energy functional $E[H]$ for our subsequent study.

\begin{center}
\begin{figure}
    \centering
    \includegraphics[width=\linewidth]{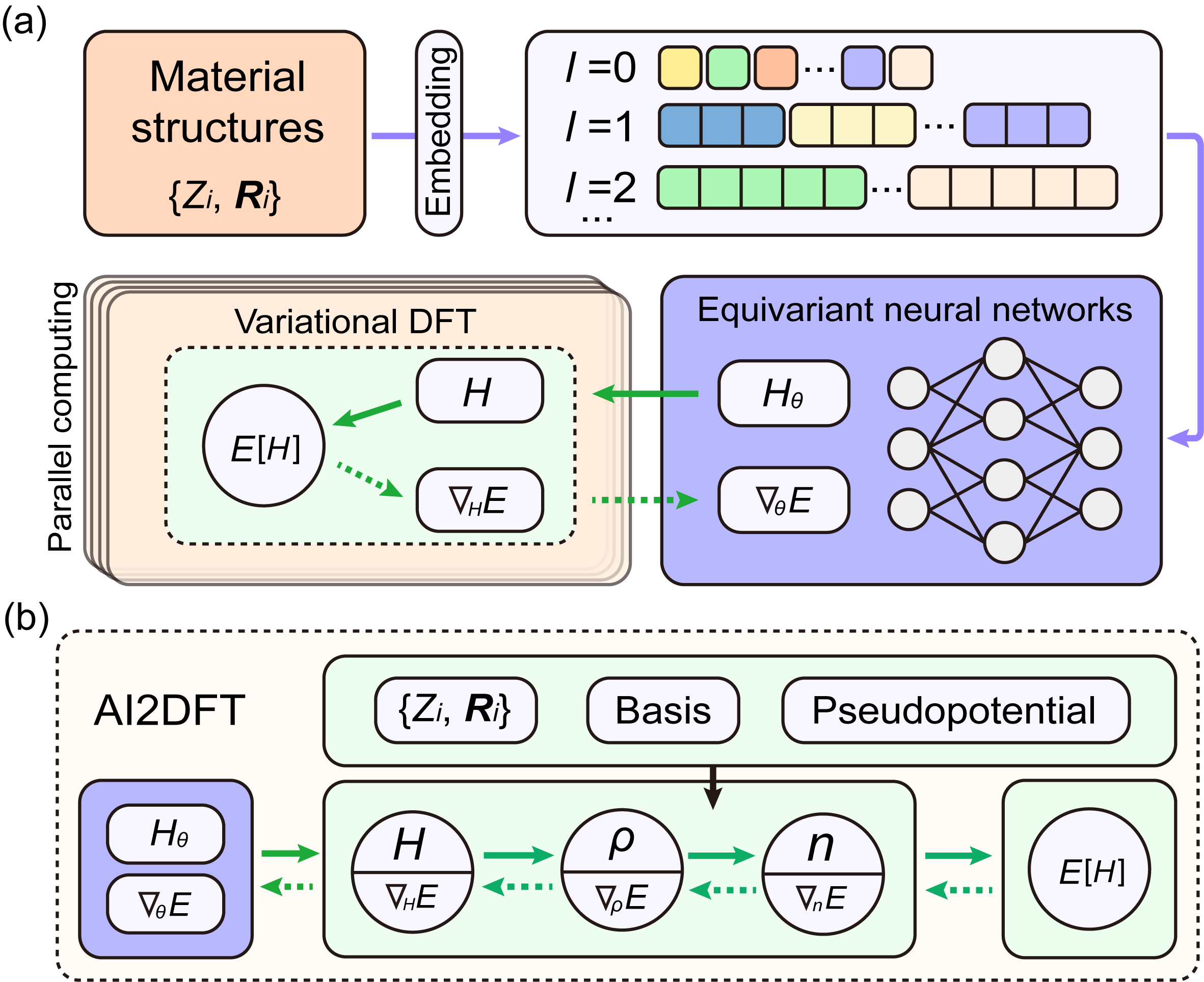}
    \caption{Architecture and implementation of neural-network DFT. (a) Overall architecture: variational DFT, which minimizes the energy functional $E[H]$, is linked to equivariant neural networks representing the DFT Hamiltonian $H_{\theta}$. $\theta$ denotes the parameters of neural networks. The structural information of materials, including the element type $Z_i$ and atomic coordinate $\boldsymbol{R}_i$ for each atom $i$, is embedded into equivariant vectors labeled by the angular momentum quantum number $l$, serving as inputs to neural networks. (b) Implementation by the AI2DFT code employing differentiable programming. The forward pass is from $H$ to density matrix $\rho$ and charge density $n$, and finally to $E[H]$. Automatic differentiation is applied to compute the gradients $\nabla_{n}E$, $\nabla_{\rho}E$, $\nabla_{H}E$, and $\nabla_{\theta}E$, and backpropagation is utilized for optimization.}
    \label{fig2}
\end{figure} 
\end{center}

Neural-network representations of the DFT Hamiltonian have been developed and applied for accelerating large-scale material simulations~\cite{deeph2022,deeph-e32023,xdeeph2023}. In the approach of deep-learning DFT Hamiltonian (DeepH)~\cite{deeph2022}, neural networks are employed to represent $H$ as a function of the material structure $\{\mathcal{R}\}$. Message-passing graph neural networks~\cite{Gilmer2017} are utilized, with vertices representing atoms and edges denoting atom pairs. Features of vertices and edges are updated through message passing from neighboring atoms, facilitating the aggregation of information from distant chemical environments. In line with the locality principle, $H$ is expressed under localized atomic basis sets. We will employ the DeepH-$E3$ framework~\cite{deeph-e32023} to ensure that the mapping $\{\mathcal{R}\}\mapsto H$ is equivariant under the Euclidean group in three-dimensional space. In DeepH-$E3$, only equivariant vectors carrying irreducible representations of the three-dimensional rotation group are permitted, and their tensor product is determined by the Wigner-Eckart theorem. The principle of equivariance is thus maintained. As illustrated in Fig.~\ref{fig2}(a), equivariant neural networks take the embedding of material structure information as input and represent the Hamiltonian matrix through their output. In this way, $H_{\theta}$ parametrized by neural-network weights is obtained, and thus the energy functional $E[H]$ can be viewed as the loss function $E[H_{\theta}]$ for neural networks.

In neural-network DFT, the DFT program must supply $\nabla_{H}E$ for optimizing neural network parameters. This poses a significant challenge in terms of DFT programming, as $\nabla_{H}E$ is not mandatory for standard SCF calculations. Fulfilling this requirement calls for a DFT program that is end-to-end differentiable. Automatic differentiation (AD) offers a suite of methods for numerically computing the derivatives of functions embedded in computer programs~\cite{ADsurvey2018}. It systematically applies the chain rule and calculus principles, eliminating the need for manual derivation, thus making it well suited for computing differential quantities like $\nabla_{H}E$. Currently, most DFT codes do not fully support the function of AD. Our neural-network DFT necessitates a differentiable implementation of DFT capable of accommodating periodic boundary conditions as well as localized atomiclike bases, a feature that, to our knowledge, has not been developed. Consequently, we have developed our own autodifferentiable DFT program named ``AI2DFT'' using the Julia language with the AD capability realized by the Zygote package~\cite{zygote2018}. Method details are described in the Supplemental Material~\cite{supp}.

Remarkably, AD is available for both DFT computation and neural-network training in AI2DFT [Fig.~\ref{fig2}(b)]. In DFT, one first derives $\rho$ and $n$ from $H$ and then uses the two quantities to compute the total energy. Based on the chain rule, AI2DFT uses the reverse-mode AD to compute $\nabla_{n}E$, $\nabla_{\rho}E$, and $\nabla_{H}E$ in turn. The gradient information $\nabla_{\theta}E$ is used in neural networks for optimization. AI2DFT supports three autodifferentiable working modes: the conventional SCF DFT and new functionalities of variational DFT and neural-network DFT.

By comparing with SCF DFT, we noticed a critical problem emerging in variational DFT: The variation of $E[H]$ can correctly predict certain ground-state properties, such as the total energy, charge density, and density matrix, but not the DFT Hamiltonian. Upon analysis of the calculated electronic structure, we observed that while the occupied manifold matches the result of SCF DFT, the unoccupied part does not. The unoccupied states have zero contribution to the ground-state energy at zero temperature, which cannot be uniquely determined via energy minimization. For computing DFT total energy, $H$ contains complete but redundant information; $\rho$ is just sufficient, which forms the foundation of density matrix functional theory~\cite{Kohn1996, DM_functional1993}. Since $\rho$ can be reliably predicted by energy minimization, one may use $\rho$ to reconstruct a new Hamiltonian, denoted as $\Tilde{H}$. The Hamiltonian reconstruction may be used for postprocessing in variational DFT, giving the ground-state Hamiltonian the same as SCF DFT.

\begin{center}
\begin{figure}
    \centering
    \includegraphics[width=\linewidth]{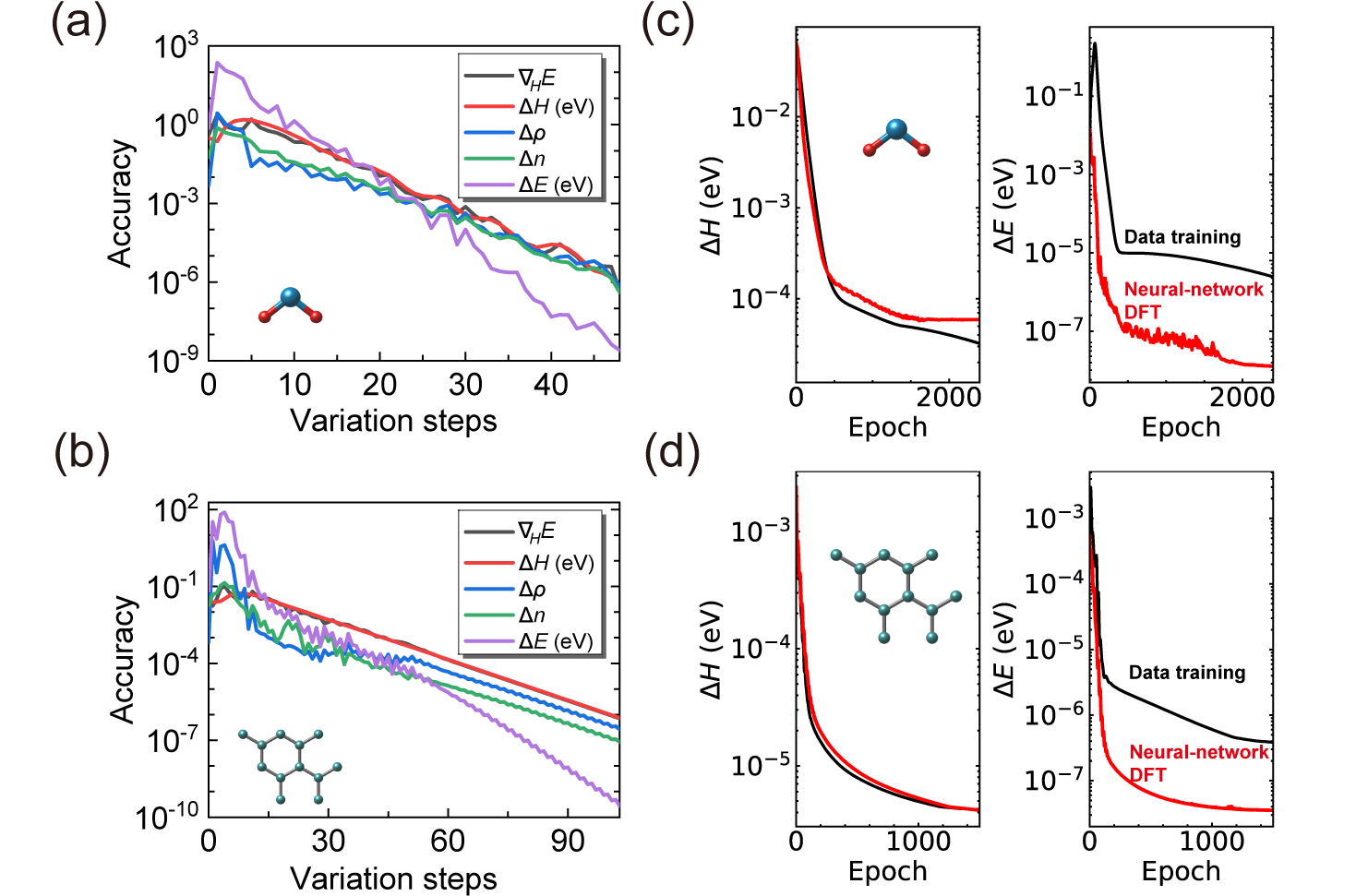}
    \caption{Validation of AI2DFT. (a,b) Variational DFT versus SCF DFT: the mean average values of $\nabla_{H}E$, $\Delta H$, $\Delta \rho$, $\Delta n$, and $\Delta E$ display exponential convergence behavior as increasing variation steps. (c),(d) Neural-network DFT versus data training: the mean average values of $\Delta H$ and $\Delta E$ steadily decrease with increasing training epochs. The upper and lower panels present example calculations on the H$_2$O molecule and graphene. Results obtained by SCF DFT calculations are used as a reference.}
    \label{fig3}
\end{figure} 
\end{center}

To be compatible with neural-network DFT, we propose to solve the above problem of the Hamiltonian in a more elegant way by introducing a generalized energy functional: $\Tilde{E}[H] = E[H] + f[H-\Tilde{H}]$, with the extra term $f[H-\Tilde{H}]$ to emulate the spirit of Hamiltonian reconstruction. The functional $f$ should attain its minimum value for $H=\Tilde{H}$, ensuring that $\Tilde{E}[H]$ shares the same ground-state energy with $E[H]$. The generalized energy gradient $\nabla_{H}\Tilde{E}$ is relevant for optimization. According to our preliminary tests, we selected a gradient of the form $\nabla_{H}E + \lambda (H-\Tilde{H})$ in our study, where $\lambda$ is an adjustable coefficient as discussed in the Supplemental Material~\cite{supp}. Note that one may replace the energy functional $E[H]$ with a free energy functional $F[H]$ for introducing temperature into DFT~\cite{MerminThermal1965}. In neural-network DFT, the second gradient term trains the Hamiltonian predicted by neural networks to resemble the reconstructed Hamiltonian. The energy gradient term $\nabla_{H}E$, which is typically not included in deep-learning DFT, drives neural networks to evolve toward lower energy configurations and enables neural networks to learn the underlying physics in an unsupervised manner. Therefore, we refer to our proposed neural-network DFT as physics-informed unsupervised learning.

We have established a theoretical framework of neural-network DFT and numerically implemented it by the AI2DFT code employing differentiable programming. Next, we comprehensively test AI2DFT by studying various types of materials, including the H$_2$O molecule, graphene, monolayer MoS$_2$, and bulk body-centered-cubic Na. These examples of materials span from molecules to periodic crystals and from metals to insulators. Firstly, we checked that SCF iterations of AI2DFT can well reproduce the benchmark results of the SIESTA code~\cite{siesta2002} (Fig. S3). Then, we applied variational DFT to study the same materials. The total energy can converge below the \(\mu\)eV scale after tens of variation steps [Figs.~\ref{fig3}(a) and \ref{fig3}(b), and Fig. S1]. Other physical quantities such as the energy gradient, Hamiltonian, density matrix, and charge density also exhibit exponential convergence behavior, validating the reliability and robustness of variational DFT.

Moreover, we assess the performance of neural-network DFT, which combines variational DFT with the DeepH-$E3$~\cite{deeph-e32023} neural networks, through comparison with conventional data-driven supervised learning approaches [Fig.~\ref{fig3}(c) and \ref{fig3}(d)]. For the study of the H$_2$O molecule, the DFT Hamiltonian can indeed be optimized to achieve a high level of accuracy: 0.06 meV by neural-network DFT and 0.02 meV by data training. Even higher accuracies of 0.004 meV are achieved for graphene using both approaches. The reliability of neural network approaches is thus confirmed. Meanwhile, we computed physical quantities using the $H$ predicted by neural networks and monitored their accuracies throughout the training process. An intriguing trend is observed: Neural-network DFT shows significantly improved performance over data training in predicting the derived physical quantities. For instance, in the study of the H$_2$O molecule, the prediction accuracy of energy reaches 0.013 \(\mu\)eV by neural-network DFT, over 60 times better than data training (0.83 \(\mu\)eV) [Fig.~\ref{fig3}(c)]. A similar trend is noticed for graphene [Fig.~\ref{fig3}(d)] as well as for other quantities such as $\rho$ and $n$ (Fig. S12). Furthermore, we devised an artificial experiment by introducing Gaussian noise to the benchmark DFT Hamiltonian to simulate numerical inaccuracies~\cite{supp}. For Hamiltonians with comparable levels of accuracy, the physical quantities obtained from neural-network DFT exhibit orders of magnitude better precision than those computed using Hamiltonians generated with Gaussian noise (Fig. S11). The underlying reason is that neural-network DFT introduces the energy gradient $\nabla_{H}E$ into optimization, enabling effective filtering of unphysical high-energy noises in the learned Hamiltonian. In this regard, the physics-informed approach surpasses conventional data-driven methods.

\begin{center}
\begin{figure}
    \centering
    \includegraphics[width=\linewidth]{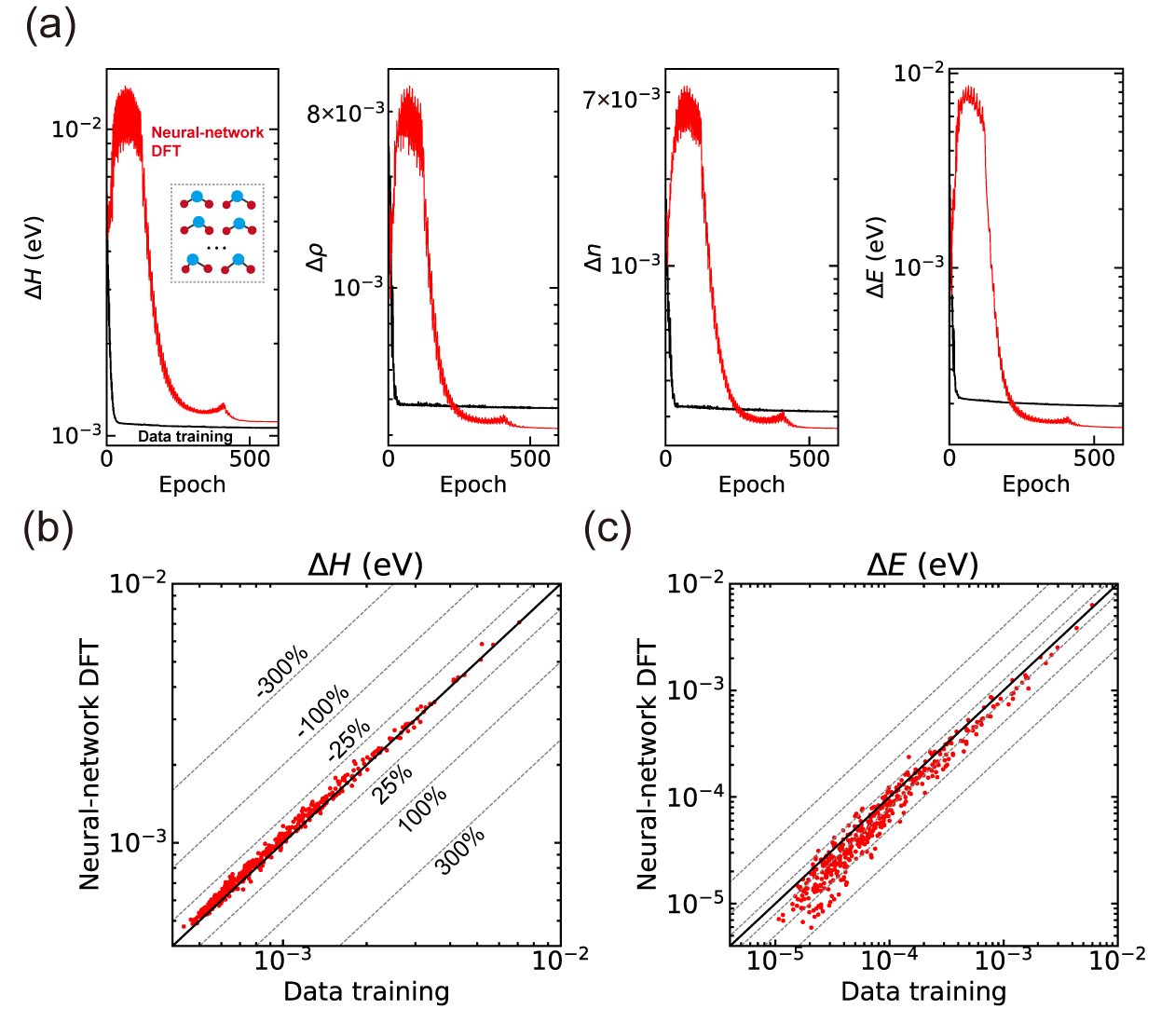}
    \caption{Unsupervised learning on multiple H$_2$O structures by neural-network DFT. (a) Learning curves of neural-network DFT compared with data-driven supervised learning. Both are initiated from a pretrained neural network model. The mean absolute errors of Hamiltonian ($\Delta H$), energy ($\Delta E$), density matrix ($\Delta \rho$), and charge density ($\Delta n$) are computed across all 300 training structures. (b),(c) Mean average errors of (b) Hamiltonian and (c) energy across 435 test structures: Neural-network DFT versus data training. The gray dashed lines indicate performance improvement ratios between the two methods.}
    \label{fig4}
\end{figure} 
\end{center}

Finally, we apply neural-network DFT to compute multiple material structures and demonstrate its capability of unsupervised learning. Using H$_2$O molecules as an example study, we initially obtained a pretrained neural network model through data-driven supervised learning using the DeepH-$E3$ approach, fine-tuned the model across 300 training structures using neural-network DFT, and achieved high prediction accuracies for Hamiltonian as well as other quantities [Fig.~\ref{fig4}(a)]. The trained neural-network model was further employed to make predictions on 435 test structures that were unseen during training, showing good generalization ability [Figs.~\ref{fig4}(b) and \ref{fig4}(c), Fig. S10, and Fig. S13] as discussed in the Supplemental Material~\cite{supp}. Compared with data-driven supervised learning, neural-network DFT yields Hamiltonians with a slightly larger mean absolute error. Nevertheless, it exhibits superior performance in predicting derived physical quantities, as confirmed by both training and test results. This is consistent with the above results, highlighting the advantage of physics-informed learning.

In summary, we proposed a theoretical framework of neural-network DFT that coherently combines variational DFT and equivariant neural networks together, enabling physics-informed unsupervised learning. The advantage of this method is elaborated upon in the Supplemental Material~\cite{supp}. Moreover, we numerically implemented neural-network DFT using deep-learning DFT Hamiltonian and differentiable programming, bringing modern techniques of automatic differentiation and backpropagation into DFT. In this context, the developments of neural networks and DFT computation are no longer isolated but will get mutual benefits and stimulate each other. The work introduces new avenues for the collaborative development of artificial intelligence and \emph{ab initio} methods, significantly enriching the scope of deep-learning \emph{ab initio} research.

\begin{acknowledgments}
We thank Xiaoxun Gong, Yuxiang Wang, and Jialin Li for useful discussions. This work was supported by the Basic Science Center Project of NSFC (Grant No. 52388201), the National Natural Science Foundation of China (Grant No. 12334003), the National Science Fund for Distinguished Young Scholars (Grant No. 12025405), the National Key Basic Research and Development Program of China (Grant No. 2023YFA1406400), the Beijing Advanced Innovation Center for Future Chip (ICFC), and the Beijing Advanced Innovation Center for Materials Genome Engineering. Yang Li is funded by the Shuimu Tsinghua Scholar program. The work was carried out at the National Supercomputer Center in Tianjin using the Tianhe new generation supercomputer.
\end{acknowledgments}

\end{document}